\documentclass[12pt,english,dvips]{scrartcl}
\usepackage{babel}
\usepackage[T1]{fontenc}
\usepackage[ansinew]{inputenc}
\usepackage{amsmath}
\usepackage{amsfonts}
\usepackage{amssymb}
\usepackage{graphicx}
\usepackage{pstricks}
\usepackage[hang]{caption2}
\usepackage{float}

\begin{document}

\begin{titlepage} \vspace{0.2in} 

\begin{center} {\LARGE \bf 

Implications of non-analytical $f(\mathcal{R})$ gravity at Solar-System scales}\\ 
\vspace*{1cm}
Orchidea Maria Lecian$^{1a}$ and Giovanni Montani$^{134b}$, \\
\vspace{0.5cm}\hfill\\

$^{1}$Dipartimento di Fisica and ICRA, Sapienza - Universit\`a di Roma,\\ P.le
Aldo Moro 5, 00185 Roma, Italy.\\
$^{3}$ENEA C.R. Frascati (Dipartimento F.P.N.), Via Enrico Fermi 45,\\
00044 Frascati, Roma, Italy.\\
$^{4}$ICRANET C. C. Pescara, \\ Piazzale della Repubblica, 10, 65100 Pescara, 
Italy.\\
$^{a}$E-mail: lecian@icra.it\\
$^{b}$E-mail: montani@icra.it
 
\vspace*{1.8cm}

\vspace*{1cm}

{\bf   Abstract  \\ } \end{center} \indent
We motivate and analyze the weak-field limit of a non-analytical Lagrangian for the gravitational field. After investigating the parameter space of the model, we impose constraints on the parameters characterizing this class of theories imposed by Solar-System data, i.e. we establish the validity range where this solution applies and refine the constraints by the comparison with planetary orbits. As a result, we claim that this class of models is viable within different astrophysical scales.  
\end{titlepage}

\section{Basic Statements}

Although General Relativity (GR) is a well settled-down theory for the description of geometrodynamics, nevertheless, over the last decades, a wide number of approaches have been developed to generalize it. These extended points of view are aimed not only at addressing the quantization of the gravitational field, but also at establishing proper deformations of the Einsteinian dynamics towards the space-time singularities.\\
From the very beginning, the possibility to reformulate GR by using a generic function of the Ricci scalar (see, for example, \cite{revi} for a recent review and the references therein) has appeared as a natural issue offered by the fundamental principles established by Einstein. Indeed, this kind of extended Einstein-Hilbert (EH) Lagrangian preserves the fundamental features at the basis of General Relativity and Equivalence principles, and induces a direct modification of the dynamics only (at least, as far as the metric approach is concerned). Recently, modified $f(\mathcal{R})$ gravity has acquired interest in view of the possibility to describe within this redefined dynamics some unexpected features observed on the large-scale structure of the Universe, i.e. the Pioneer anomaly, the galaxy rotation curve behaviour, the Universe acceleration, and, finally, the removal of singular behaviours of the gravitational field \cite{var}.\\
However, it is important to remark that any modification of the EH Lagrangian is reflected onto a deformed gravitational-field dynamics at any length scale investigated or observed. Thus, the success of such $f(\mathcal{R})$ gravity in the solution of a specific problem has to match consistency with observation in other length scales \cite{zac,will,tak}. In this respect, we emphasize that the small value of the present curvature of the Universe \cite{wmapcurv} leads us to believe that, independently of its specific functional form, the $f(\mathcal{R})$ term must be regarded as a lower-order expansion in the Ricci scalar. On the other hand, it is easily understood that the peculiarities of such an expansion will be extremely sensitive of the morphology of the deformed Lagrangian.\\
The most immediate generalization is of course to deal with a function of the Ricci scalar analytical in the point $R=0$, so that its Taylor expansion holds \cite{cap}. This approach is equivalent to deal with a polynomial form \cite{large,stelle}, whose free parameters are available to fit the observed phenomena on different sectors of investigation. Despite the appealing profile of such a choice, it is extremely important to observe that it is not the most general case, since real (non-integer) exponents of the Ricci scalar are in principle on the same footing as the simplest case \cite{sc1,nonrat}. In this letter, we will concentrate our attention on such an open issue, and we will develop a modified theory of the form $f(\mathcal{R})= \mathcal{R}+\gamma \mathcal{R}^\beta$, where $\gamma$ and $\beta$ are two free parameters to be constrained at physical level \cite{tak,zac}. In particular, the details of our model will lead us to deal with rational non-integer numbers for $\beta$, and to restrict it in the most appropriate interval for physical interpretation at low curvature, $2<\beta<3$.\\
We analyze the weak-field limit of our modified Lagrangian and derive the corresponding spherically-symmetric field equations. Here we retain only those non-integer powers in the Ricci scalar, which represent the dominant effect after the linear approximation. The choice $2<\beta<3$ allows us to distinguish the dominant non-Einsteinian terms from the non-linear ones of GR. The explicit solution of the derived system is found, and its main features are analyzed in view of possible constraints from observational data\cite{will,tak,gil}.  
The main issue of our treatment is to demonstrate that our $f(\mathcal{R})$ model is appropriate both to fulfill Solar-System constraints and to provide a significant break-down of the Newton law at galactic scales. The paper is organized as follows. In section two, we provide the details of the model. In section three, we analyze the weak-field limit of the $f(\mathcal{R})$ theory. Section four is eventually devoted to phenomenological estimations, which fix the constraints on the model. Discussion and conclusion follow.
   
\section{Non-analytical $f(\mathcal{R})$ model}
We consider the following modified gravitational action in the Jordan frame
\begin{equation}\label{nonan}
S=-\tfrac{c^3}{16\pi G}\int d^3 x \sqrt{-g} f(\mathcal{R})=-\tfrac{c^3}{16\pi G}\int d^3 x \sqrt{-g} \left(\mathcal{R}+\gamma \mathcal{R}^\beta \right),
\end{equation}
where $\beta$ is dimensionless, and the parameter $\gamma$ has the dimensions of {\it length} $^{2\beta-2}$. We can define the characteristic length scale of our model as $L_\gamma\equiv\mid \gamma \mid ^{1/(2\beta-2)}$. It is straightforward to verify that (\ref{nonan}) is non-analytical in $\mathcal{R}=0$ for non-integer, rational $\beta$, i.e. its Taylor expansion in the vicinity of $\mathcal{R}=0$ does not hold. Similar models have been investigated, for example, in \cite{sc1}, mostly from a cosmological point of view; in \cite{nonrat}, the possibility of an irrational exponent is also envisaged.\\
We can gain further information on the value of $\beta$ by analyzing the conditions that allow for a consistent weak-field stationary limit. Furthermore, the study of the potential of the scalar field in the Einstein frame will define the whole parameter space $\{\beta\gamma\}$. Having in mind to investigate the weak field limit of our theory to pursue its prediction at Solar-System scales, we can decompose the corresponding metric as $g_{\mu\nu}\equiv\eta_{\mu\nu}+ h_{\mu\nu}$, where $h_{\mu\nu}$ is a small (for our case, static) perturbation of the Minkowskian metric $\eta_{\mu\nu}$. In this limit, the Einstein equations read
\begin{subequations}
\begin{align}
R_{\mu\nu}-\tfrac{1}{2}\eta_{\mu\nu}R-\Gamma\beta R^{\beta-1};_{\mu\nu}+\gamma\beta\eta_{\mu\nu}\Box R^{\beta-1}=0\label{mfe1}\\
R=3\Gamma\beta\Box R^{\beta-1}\label{mfe2}
\end{align}
\end {subequations}
where $;$ denotes covariant differentiation, $\Box\equiv g^{\rho\sigma}\nabla_\rho\nabla_\sigma$, and, for the sake of compactness, in this treatment we retain the usual notation of the curvature scalar $R$ and the Ricci tensor $R_{\mu\nu}$ calculated up to the weak-field limit, as we are evaluating the $\mathcal{O}(h)$ orders.\\The structure of such field equations lead us to focus our attention on the restricted region of the parameter space $2<\beta<3$. This choice is enforced by the fulfillment of the conditions by which all other terms are negligible with respect to the linear and the lowest-order non-Einsteinian ones.\\ 
The modified gravitational action (\ref{nonan}) can be cast in an equivalent scalar-tensor model in the Einstein frame \cite{nonrat,thom}, i.e. a scalar field $\varphi$ minimally coupled to gravity, by means of the conformal transformation $g_{\mu\nu}\rightarrow e^{\varphi}g_{\mu\nu}$, where $\varphi\equiv-\ln f'$. In addition, the further transformation $\varphi\rightarrow k\phi$, where $k\equiv\sqrt\frac{16\pi G}{3c^4}$ accounts for the right dimensions of a scalar field. Within this framework, the potential $V(\varphi)$ describing the dynamics of the scalar field reads 
\begin{equation}\label{scalpot}
V(\varphi)=e^{2k\,\phi }\,\left( -1 + \beta  \right) \,
  {\left( \tfrac{-1 + e^{-k\varphi }}{\beta \,\gamma } \right) }^
   {\tfrac{\beta }{-1 + \beta }}\,\gamma
\end{equation}
The appearance of a minimum in the potential is crucial, since the cosmological implementation of this picture into an isotropic and homogeneous Universe suggests one that such a minimum becomes, sooner or later, an attractive stable configuration for the system \footnote{ In fact the energy density of the scalar field is monotonically damped by the Universe expansion and the later stages of the Universe must be compatible with a weak perturbation of the GR scheme. As a matter of fact, a constant value of $\varphi$ corresponds to an Einsteinian dynamics and its extremal-point character ensures the nearby stability of this configuration.}.\\
From the analysis of the scalar potential, we see that, for $2<\beta<3$, there are two extremal points
\begin{equation}\nonumber
\varphi_0=0,\ \ {\varphi_\beta = 
   \tfrac{1}{k}{\log \left(\tfrac{-2 + \beta }{2\,\left( -1 + \beta  \right) }\right)}}
\end{equation}
and $V(\varphi)$ behaves likes an attractor only for the cases
\begin{enumerate}
  \item\label{uno} $\gamma>0$, $\beta>2$, $\beta=\tfrac{2n}{2m+1}$, $m,n\in\mathbb{Z}$: $\varphi_0$ is a minimum, $\varphi_\beta$ is a maximum;
	\item\label{due} $\gamma<0$, $\beta>2$, $\beta=\tfrac{2n}{2m+1}$, $m,n\in\mathbb{Z}$: $\varphi_0$ is a maximum, $\varphi_\beta$ is a minimum.
\end{enumerate}

\section{Weak-field limit}
From the analysis of the form of (\ref{mfe1}-\ref{mfe2}) in the Jordan frame, we learn that it is possible to find a post-Newtonian solution by solving the set of (\ref{mfe1}-\ref{mfe2}) up to next-to-leading order in $h$, i.e. up to $\mathcal{O}(h^{\beta-1})$, and neglecting the $\mathcal{O}(h^2)$ contributions only for the cases $2<\beta<3$.\\
These considerations motivate the choice we claimed above concerning the restriction of the parameter $\beta\in(2,3)$.
The most general spherically-symmetric line element in the weak-field approximation reads
\begin{equation}\label{line}
ds^2=(1+\Phi)dt^2-(1-\Psi)dr^2-d\Omega^2,
\end{equation}
where $\Phi$ and $\Psi$ are the two generalized gravitational potentials, and $d\Omega^2$ is the solid-angle element. Within this framework, the modified Einstein equations (\ref{mfe1}-\ref{mfe2}) rewrite 
\begin{subequations}
\begin{align}
R_{tt}-\tfrac{1}{2}R-\gamma\beta\nabla^2 R^{\beta-1}=0\label{wfme1}\\
R_{rr}+\tfrac{1}{2}R-\Gamma\beta R^{\beta-1},_{rr}+\gamma\beta\nabla^2 R^{\beta-1}=0\label{wfme2}\\
R_{\theta\theta}+\tfrac{1}{2}r^2R-\Gamma\beta rR^{\beta-1},_{r}+\gamma\beta r^2\nabla^2 R^{\beta-1}=0\label{theta}\\
R=-3\Gamma\beta\nabla^2 R^{\beta-1},\label{trace}
\end{align}
\end{subequations}
where
\begin{subequations}
\begin{align}
R=\nabla^2\Phi+\tfrac{2}{r^2}(r\Psi),_r\label{ricci}\\
R_{tt}=\tfrac{1}{2}\nabla^2\Phi\\
R_{rr}=-\tfrac{1}{2}\Phi,_{rr}-\tfrac{1}{r}\Psi,_{r}\\
R_{\theta\theta}=-\Psi-\tfrac{r}{2}\Phi,_r-\tfrac{r}{2}\Psi,_r\\
R_{\phi\phi}=\sin ^2 \theta R_{\theta\theta}.\label{phiphi}
\end{align}
\end{subequations}
Here $,$ denotes ordinary differentiation and $\nabla^2\equiv \tfrac{d^2}{dr^2}+\tfrac{2}{r}\tfrac{d}{dr}$.\\
System (\ref{wfme1}-\ref{trace}) is solved by
\begin{subequations}
\begin{align} 
R=Ar^{\tfrac{2}{\beta-2}}, \ \ A=\left[-\tfrac{6\gamma\beta(3\beta-4)(\beta-1)}{(\beta-2)^2}\right]^{\tfrac{1}{2-\beta}}\label{pns1}\\
\Phi=\sigma+\tfrac{\delta}{r}+\Phi_\beta\left(\tfrac{r}{L_\gamma}\right)^{2\tfrac{\beta-1}{\beta-2}},\ \ \Phi_\beta\equiv\left[-\tfrac{6\beta(3\beta-4)(\beta-1)}{(\beta-2)^2}\right]^{\tfrac{1}{2-\beta}}\tfrac{(\beta-2)^2}{6(3\beta-4)(\beta-1)}\label{pns2}\\
\Psi=\tfrac{\delta}{r}+\Psi_\beta \left(\tfrac{r}{L_\gamma}\right)^{2\tfrac{\beta-1}{\beta-2}},\ \ \Psi_\beta\equiv\left[-\tfrac{6\beta(3\beta-4)(\beta-1)}{(\beta-2)^2}\right]^{\tfrac{1}{2-\beta}}\tfrac{(\beta-2)}{3(3\beta-4)}\label{pns3} 
\end{align}
\end{subequations}
where the integration constant $\delta$ has the dimensions of {\it length}, and the dimensionless integration constant $\sigma$ can be set equal to zero. The integration constant $A$ has the dimensions of {\it length}$^{\tfrac{2\beta-2}{2-\beta}}$, and $\Phi_\beta$ and $\Psi_\beta$ are dimensionless, accordingly. Moreover, we find that $A$ is well-defined only in the case $\gamma<0$, $\beta=2n/(2m+1)$, i.e. only for configuration (\ref{due}) of the scalar-tensor description. In this case, $A>0$. It is remarkable that the limit $\gamma\rightarrow0$, apparently restoring GR \cite{sot1}, is mapped, in the solution, into the opposite case $\gamma\rightarrow\infty$ \cite{stelle}. Indeed, this situation in which the typical length scale associated to the parameter $\gamma$ is arbitrarily large would correspond to the physical intuition of an arbitrarily large scale for the relevance of the non-Einsteinian contributions.

\section{Solar-System constraints}
The most suitable arena where to evaluate the reliability and the validity range of the weak-field solution (\ref{pns1}-\ref{pns3}) is, of course, the Solar System \cite{tak,zac}. For this reason, we can specify (\ref{pns2}) and (\ref{pns3})for the typical length scales involved in the problem. To this end, we split $\Phi$ and $\Psi$ of (\ref{pns2}) and (\ref{pns3}) into two terms, the Newtonian part and the post-Newtonian one, respectively, i.e.
$\Phi\equiv\Phi_N+\Phi_{PN}\equiv\-r_s/r+ \Phi_\beta(r/L_\gamma)^{2(\beta-1)(\beta-2)}$ and 
$\Psi\equiv\Psi_N+\Psi_{PN}\equiv-r_s/r+ \Psi_\beta(r/L_\gamma)^{2(\beta-1)(\beta-2)}$,
where the integration constant $\delta$ in (\ref{pns2}-\ref{pns3}) results naturally as the Schwarzschild radius of the Sun (apart from the sign) $r_s\equiv2GM_s/c^2$, $M_s$ being the Solar mass. In fact, while the weak-field approximation of the Schwarzschild metric is valid within the range $r_s<<r<\infty$ because it is asymptotically flat, the post-Newtonian correction has the peculiar feature to diverge for $r\rightarrow\infty$. It is therefore necessary to establish a validity range $r_{min}<<r<<r_{max}$, where this solution is physically predictive\cite{extmetric}. The definition of this validity range is strictly related to the parameters $\beta$ and $L_\gamma$ involved in the model. Stringent constraints for $L_\gamma$ can be found in the analysis of planetary orbital periods, which represent a severe test within the Solar System, because of the high precision of experimental data.  
\subsection{Validity range}
Since we aim to provide a physical picture at least of the planetary region of the Solar System, we are led to require that $\Phi_{PN}$ and $\Psi_{PN}$ remain small perturbations with respect to $\Phi_N$ and $\Psi_N$, so that it is easy to recognize the absence of a minimal radius except for the condition $r>>r_s$. A maximum radius $r_{max}$ appears instead.
In fact, the definition of $r_{max}$ requires further discussion. The typical distance $r^*$ corresponds to the request 
\begin{equation}\label{rmax}
\mid\Phi_N(r^*)\mid\sim\Phi_{PN}(r^*),\ \ \mid\Psi_N(r^*)\mid\sim\Psi_{PN}(r^*).
\end{equation}
For $r_s<<r<<r^*$, the system obeys thus Newtonian physics, and experiences the post-Newtonian term as a correction. Another maximum distance $r^{**}$ can be defined, according to the request that the weak-field expansion in (\ref{line}) should hold, regardless to the ratios $\Phi_{PN}/\Phi_N$ and $\Psi_{PN}/\Psi_N$. According to our definition of $r_{min}$, $r^{**}$ is defined as
\begin{equation}\label{star}
\mid\Phi_N(r^{**})\mid<<\Phi_{PN}(r^{**})\sim1,\ \ \mid\Psi_N(r^{**})\mid<<\Psi_{PN}(r^{**})\sim1.
\end{equation}
We remark that that, within this scheme, $r^*$ and $r^{**}$ are defined as functions of $\beta$ and $L_\gamma$, i.e. $r^*\equiv r^*(\beta, L_\gamma)$ and $r^{**}\equiv r^{**}(\beta, L_\gamma)$ respectively
\begin{equation}
r^*\sim\left(r_s/\Phi_\beta\right)^{\tfrac{\beta-2}{3\beta-4}}L_\gamma^{\tfrac{2\beta-2}{3\beta-4}},\ \ r^{**}\sim L_\gamma\big/\Phi_\beta^{\tfrac{\beta-2}{2\beta-2}}.
\end{equation}
It is important to stress that, for the validity of our scheme, the condition $r^*>>r_s$ must hold, i.e. $L_\gamma>>\Phi_\beta^{(\beta-2)/(2\beta-2)}r_s$.  
\subsection{Planetary orbital periods}
At this level, neglecting the lower-order effects concerning the eccentricity of the planetary orbit, we can deal with the simple model of a planet moving on circular orbit around the Sun, and its orbital period $T$ is given by $T=2\pi (r/a)^{1/2}$,  $a\equiv\tfrac{c^2}{2}d\Phi/dr$ being the centripetal acceleration. For our model, from (\ref{pns2}), we get
\begin{equation}\label{tibeta}
T_\beta=\tfrac{2\pi}{(GM_s)^{1/2}}r^{3/2}\left(1+2\tfrac{\beta-1}{\beta-2}\Phi_\beta\tfrac{r^{\tfrac{3\beta-4}{\beta-2}}}{r_sL_\gamma^{\tfrac{2\beta-2}{\beta-2}}}\right)^{-1/2}.
\end{equation} 
Hence we evaluate the correction to the Keplerian period $T_K=2\pi r^{3/2}(GM_s)^{-1/2}$, compare it with he experimental data of the period $T_{exp}$ and its uncertainty $\delta T_{exp}$ and then impose that the correction be smaller than the experimental uncertainty, i.e.
\begin{equation}\label{last}
\tfrac{\delta T_{exp}}{T_{exp}}\geq\tfrac{\mid T_K-T_\beta\mid}{T_K}\sim\tfrac{\beta-1}{\beta-2}\Phi_\beta\tfrac{r_p^{\tfrac{3\beta-4}{\beta-2}}}{r_sL_\gamma^{\tfrac{2\beta-2}{\beta-2}}},
\end{equation}
where $r_p$ is the mean orbital distance of a given planet from the Sun.
\subsection{Numerical estimations}
We now specify the previous considerations for and intermediate value of $\beta$, allowed by the analysis of the weak-field limit consistency and the scalar-tensor description. To provide proper numbers, we choose a typical (non-peculiar) value of the parameter $\beta$, say $\beta=8/3$.\\
High-precision measurements are nowadays available for the distances between Solar-System planets and the Sun, so that the relative error in the orbital period is extremely small. According to this fact, we specify our analysis for example for the Earth \cite{zac}, for which $T_{exp} = 365.256363051 days$, $\delta T_{exp}=5.0\cdot10^{-10} days$ and $r_p=149.6\cdot10^6 km$. For $\beta=8/3$, (\ref{last}) implies $L_\gamma>1.147466382\cdot 10^{11} Km$.\\
This lower bound for $L_\gamma$ can be plugged into (\ref{rmax}) and (\ref{star}) to get an estimation of the distances $r^*$ and $r^{**}$. Since the constraints on the $tt$ and the $rr$ components of the metric tensor imply similar conditions, we restrict our analysis to the physically-relevant case of the $tt$ component, and, after direct calculation, obtain $r^*\sim1.6\cdot10^{10} km$ and $r^{**}\sim1.5\cdot10^{12} km$.\\ 
Such values of $r^*$, $r^{**}$ and $L_\gamma$ are essentially stable with respect to the range of $\tfrac{8}{3}<<\beta<<3$. In fact, in the vicinity of the greatest value $\beta=3$, we would obtain $L_\gamma>1.3\cdot10^{12} km$, $r^*\sim4\cdot10^{10} km$, $r^{**}\sim1.3\cdot10^{13} km$. Our analysis clarifies how the predictions of the corresponding equations for the weak-field limit appear viable in view of the constraints arising from the Solar-System physics. Indeed, the lower bound for $L_\gamma$ does not represent a serious shortcoming of the model, as we are going to discuss.

\section{Discussion and Conclusions}
When adopting an $f(\mathcal{R})$ model to extend GR, we are addressing an appealing procedure to consistently derive unexpected features of standard geometrodynamics. However, the wide spectrum of possible effects allowed by such a reformulation could represent one of the shortcomings of the approach if no precise account of observational data is taken. Indeed, we have enough precision in the measurement of astrophysical systems to represent a severe test for the reliability of an extended approach for the gravitational-field dynamics. The present investigation has demonstrated how, in contrast with other analogous treatments\cite{erreenne,zac}, the non-analytical case $2<\beta<3$ is not excluded by Solar-System tests. Even though we have restricted our analysis here to the uncertainty of the Earth period, nevertheless in sheds light on the viability of the theory unambiguously. The lower bound of $L_\gamma$ is almost compatible already with the Solar-System scale, since it would predict non-Newtonian effects for outer regions only. However, it is clear that the corresponding value of $r^{**}$, where the theory would require a non-linear treatment, is manifestly incompatible with observations on the galactic scale. Thus, we can obtain a more reliable estimation for the fundamental scale $L_\gamma$ by requiring that the value of $r^*$ overlap the typical galactic scale on which the rotation curves manifest the flatness of their behaviour, as outlined by $21 cm$ Hydrogen line \cite{21}. Indeed, taking $r^*$ in correspondence of the length of about $10 Kpc\sim3\cdot10^{17} km$ and an internal mass (assumed as spherically distributed) of $10^{11} M_s$, we get $L_\gamma\sim4\cdot10^{17} km$ and $r^{**}\sim5\cdot10^{18}km$ for $\beta=8/3$. It is not among the purposes of this short communication to discuss the detailed predictions of our theories on the galactic scales (we will address the analysis of this question in further investigations). What we wish to emphasize here is that our proposal for a suitable value of the characteristic length $L_\gamma$ can predict small changes on Solar-System observables, while it could drive relevant modifications in higher-scale astrophysics.

\end{document}